\documentclass{aa}  

\usepackage[utf8]{inputenc}
\usepackage{graphicx}
\usepackage{txfonts}
\usepackage{xcolor}

\usepackage[colorlinks,linkcolor=blue,citecolor=blue,urlcolor=blue]{hyperref}
\usepackage{orcidlink}

\begin{document}
\title{Automated detection of exploding granules with SDO/HMI data}

\author{
J. Ballot\orcidlink{0000-0002-9649-1013} 
\and
T. Roudier\orcidlink{0000-0003-2071-7511}
}

\institute{
Institut de Recherche en Astrophysique et Planétologie (IRAP), 
Université de Toulouse, CNRS, UPS, CNES, 
14 avenue Edouard Belin, 31400 Toulouse, France\\
\email{jerome.ballot@irap.omp.eu}
}

\date{Received 15 June 2024 / Accepted 5 September 2024}
\authorrunning{J. Ballot \& T. Roudier}

\abstract
 {Exploding granules on the solar surface play a major role in the dynamics of the outer part of the convection zone, especially in the diffusion of the magnetic field.}
{We aim to develop an automated procedure able to investigate the location and evolution of exploding granules over the solar surface and to get rid of visual detection.}
  {We used sequences of observations of intensity and Doppler velocity, as well as magnetograms, provided by the Helioseismic and Magnetic Imager aboard the Solar Dynamics Observatory. 
  The automated detection of the exploding granules was performed by applying criteria on either three or two parameters: the granule area, the amplitude of the velocity field divergence, and, at the disc centre, the radial Doppler velocity. Our analyses show that granule area and divergence amplitudes are sufficient to detect the largest exploding granules; thus, we can automatically detect them, not only at the disc centre, but across the whole solar surface.
  }
  {Using a 24-hour-long observation sequence, we have demonstrated the important contribution of the most dynamic exploding granules in the diffusion of the magnetic field in the quiet Sun. Indeed, we have shown that the most intense exploding granules are sufficient to build a large part of the photospheric network. We have also applied our procedure on \textit{Hinode} observations to locate the exploding granules relative to trees of fragmenting granules (TFGs). We conclude that, during a first phase of about 300 minutes after the birth of a TFG, exploding granules are preferentially located on its edge. Finally, we also show that the distribution
  of exploding granules is homogeneous (at the level of our measurement errors) over the solar surface without a significant dependency on latitude.}
 {}

\keywords{Sun: granulation -- Sun: atmosphere}

\maketitle
\nolinenumbers
\section{Introduction}

The entire surface of the Sun, except in sunspot regions, is renewed every 15 to 20 minutes by the motions of turbulent convection. The solar granulation represents the smallest visible scales of these motions at the top of the convective layer and contributes to this change on very short timescales. The distribution of the observed solar granules lies between 58 km, the limit of the best spatial resolution today \citep[DKIST,][]{Rimmele2020}, and 3000 km \citep{Roudier1986}.
Some granules rapidly increase in size and a dark spot appears in their centre before they split into several pieces. Some fragments become proper granules, while the others vanish. These quickly expanding granules have been called ‘exploding granules’ (or sometimes ‘exploders’) and were discovered and first characterised in the 1950s and 1960s by J.~Rösch and colleagues thanks to high-resolution cinematographic pictures of the solar surface taken at the Pic du Midi Observatory \citep[see][]{Carlier68}. One generally names the moment when they break an `explosion', even if it is not a physical explosion.
The exploding granules
are the most vigorous elements of the superficial convective motions and have a large influence on the dynamics of the photosphere, in particular on the diffusion of magnetic field in the quiet Sun \citep{Roudier2020} or on the formation of the magnetic network \citep{Roudier2003,Roudier2016,Malherbe2018}, while they cover only 2.5\% of the photosphere \citep{Namba86}. The break-up of an exploding granule influences the movements of its neighbours up to a radius of around $4\arcsec$ (2.9~Mm) and produces a divergence signal on spatial and temporal mesogranular scales  \citep{Delmoro2004}.

 The solar granulation is organised into families of granules forming the so-called trees of fragmenting granules (TFGs), which can last for several hours \citep{Roudier2003}. The TFGs are formed by successively exploding granules originating from a single parent. They contribute to the network formation in the quiet Sun by efficiently diffusing the magnetic field \citep{Roudier2016}. 
 A causal link between the supergranulation and exploding granules is also invoked by \citet{Rast1995}. This author concluded that the supergranulation is related to the spatial distribution of exploding granules and is a secondary manifestation of granulation itself. 

 Unfortunately, the limited ($0.5\arcsec$ pixel) resolution of the images provided by the Helioseismic and Magnetic Imager (HMI) on board the Solar Dynamics Observatory (SDO) relative to those of the \textit{Hinode} satellite, which is five times larger, prevents us from detecting entire families with the SDO/HMI data. 
 However, the detection of exploding granules is possible as it has been demonstrated in \citet{Roudier2020} by using de-convolved SDO/HMI images. In this study, they evaluate the detection and the expansion of exploding granules at several wavelengths, and at different spatial and temporal resolutions from ground-based and space-borne observatories (including SDO/HMI), and recover occurrence rates of exploding granules similar to those found in  previous observations \citep{Palacios2012}.

The main goal of the present study is to take advantage of the SDO/HMI observations to automatically detect exploding granules across the whole solar surface. 
We investigate a way to detect automatically the exploding granules. In Sect.~\ref{sec:red}, we describe the data selection and our reduction pipeline to get the velocity field in spherical co-ordinates and morphological parameters. In Sect.~\ref{sec:detect_meth}, we present in more detail the method developed to identify exploding granules by using their area, the divergence field,
and the vertical velocity field when it is available.
We describe the location of these exploding granules relative to the TFGs in Sect.~\ref{sec:TFG}. The detection of the exploding granules, and their actions on the magnetic field, at different latitudes and longitudes (along the equator) is presented in Sect. \ref{sec:lat}. Finally,  we conclude in Sect.~\ref{sec:conc} by pointing out  the importance of such exploding granules for the dynamic of the surface flow in diffusing the magnetic field over the solar surface.

\section{SDO/HMI data selection and reduction}\label{sec:red}

HMI \citep{Scherrer2012,Schou2012} aboard the SDO provides uninterrupted observations of the entire solar disc. This allows us to extract a uniform sequence of intensity images covering a whole day. We used the SDO/HMI white-light data from 12:00 U.T. on 5 April 2019 to 12:00 U.T. on 6 April 2019 with a time step of 45\:s. The pixel scale is about 0.5\arcsec, corresponding to 365.8\:km at the solar surface. The exact values of pixel scale have been used to reduce and analyse each image. The conversion from arcseconds to kilometres properly accounts for the exact distance between the SDO satellite and the Sun.

Different corrections have been applied in alignment \citep{Rincon2017}.
The differential rotation profile, measured from the raw HMI Doppler images and averaged over 24h, was used to remove the mean horizontal flow, following the ‘derotation’ procedure described in \citet{Rincon2017}.
To summarise, it consists of correcting the mean differential rotation of the Sun to bring back 
regions of the solar surface at the location they hold at a reference time.
In the present study, the reference for the derotation was taken at 12:00 U.T. on 5 April 2019, which is the first image of the sequence. Thus, we can follow the evolution in time of the structures at their original longitude.

White-light intensity $4096\times4096$ images were de-convolved from the HMI transfer function \citep{Couv2016} and re-binned by a factor of two, resulting in images with a size of $8192\times 8192$ and a pixel size of about $0.25\arcsec$ (182.9\:km). This operation allows us to increase the number of pixels for each granule and improve the segmentation processing. Such a technique has already been successfully applied in \citet{Roudier2020}. More precisely, the most efficient segmentation algorithm for solar granulation in our case is a segmentation curvature-based criterion described
in \citet{Rieutord2007}. This image segmentation is very efficient at removing large-scale intensity fluctuations.
 To limit projection effects, the field of view is restricted to $250\times250$ pixels. At the disc centre, it corresponds to about $62.5\arcsec\times62.5\arcsec$, with a slight dependence on the distance to the Sun during the orbit. The tracked data cubes (intensity and Doppler) were filtered in the $k-\omega$ domain with a threshold phase velocity of 6~$\mathrm{km\, s^{-1}}$ to keep only convective motions 
 and remove acoustic waves
 \citep{Title1989}.

For comparison purposes, in Sect.~\ref{sec:TFG}, we also used a time sequence of solar granulation observations between 08:04~U.T. and 10:59~U.T. on 30 August 2010 (1h55m in length), which were obtained simultaneously with the \textit{Hinode} and SDO satellites. They have a pixel size of  $0.11\arcsec$ and $0.5\arcsec$, respectively. Both sequences were re-binned with an equivalent pixel size of 0.25\arcsec.  These sequences  were aligned, filtered for 5-minute oscillations, and then segmented. Finally, we also used a 48-h sequence of granulation obtained by \textit{Hinode} from 29 to 31 August 2007 already reduced by \citet{Roudier2016} to locate exploding granules relative to the TFGs in Sect.~\ref{sec:TFG}.

\section{Exploding granule detection}\label{sec:detect_meth}

The detection of exploding granules by eye used in the previous works \citep{Roudier2020} is limited, since it is time-consuming and tedious. 
Moreover, the very large simultaneous number of exploding granules at a given time makes the task more difficult because the eye jumps from one granule to another constantly. Thus, the required concentration is greatly multiplied.  We can add that a bias in the selection of the exploding granules acts despite the observer, who will tend to select only the best-shaped exploding granules close to the circular shape. 
Conversely, an automated detection avoids such biases but requires a quantitative definition of selection parameters.
We thus identify exploding granules as the granules (i) with the largest area, (ii) located where the divergence of the horizontal velocity flows is the largest (at the spatial scale of granules), and (iii) with a large ascending velocity to select the most powerful convective elements.
The granule areas were measured by isolating the granules thanks to an image segmentation method \citep[e.g.][]{Rieutord2007}.
The determination of the horizontal flows was performed from intensity images of HMI by using the local correlation technique \citep[LCT,][]{November1988}. We have access to the vertical velocity thanks to Doppler velocity measurements, filtered from the acoustic oscillations,  provided by HMI at the same time as the intensity images. The vertical velocity is thus only available close to the disc centre.

\subsection{Determining selection criteria}\label{thres}
 
In this section, we quantitatively define the three selection criteria on the granule area, the horizontal flow divergence, and the vertical velocity, by giving the threshold they must exceed to locate exploding granules.

First, we retained only the largest granules. We chose as a threshold a minimum area of $1\:\mathrm{Mm^{2}}$ -- that is, 30 pixels -- because it corresponds to the mean area reached by the granules during their evolution, measured by \citet{Hirz1999}: they showed that the distribution of maximal area reached by a granule during its evolution lie between $0.8\:\mathrm{Mm^{2}}$ and $1.3\:\mathrm{Mm^{2}}$ and peaks around $1\:\mathrm{Mm^{2}}$.
These large granules occupy 50\% of the area covered by granules at the solar surface, but they represent less than 25\% of the total number of granules  \citep{Roudier1986}. In this way, the smaller exploding granules are lost, but our main goal is to study the flow field generated by the large ones and their effects on the diffusion of the magnetic field in the quiet Sun. We also tested a smaller threshold of 20 pixels ($0.66\:\mathrm{Mm^{2}}$), which allowed us to keep smaller exploding granules but increase the level of false detection (see further below).

Second, the divergence amplitude threshold must depend on the spatial and temporal windows used in the LCT to calculate horizontal velocities. The amplitude of velocities and divergences deduced with LCT depend on the spatial window \citep{Verma2011} and temporal window \citep{SBN94}.
For example, \citet{Strous1996} found that the divergence had a maximum amplitude of $1.310\cdot 10^{-4}\:\mathrm{s^{-1}}$ when they used the LCT with a spatial window of 7.6\:Mm and a temporal window of 92 min. On their side, \citet{Potzi2005} found a positive divergence amplitude of up to $5\cdot 10^{-4}\:\mathrm{s^{-1}}$ with a spatial window of 3.5\:Mm and a time window of 100 min, while 
\citet{Verma2011} found maximum values for divergence of $1.5\cdot 10^{-3}\:\mathrm{s^{-1}}$ with a spatial window of  2.56\:Mm and a temporal window of 60\:min. 
Variations between these different studies reach an order of magnitude.
These differences are easy to explain, since higher-resolution maps capture more small-scale motions. Reassuringly, the 10th percentile values for divergence are essentially the same  ($\sim 4\cdot 10^{-4}\:\mathrm{s^{-1}}$) for different studies \citep{SBN94,Verma2011}. Moreover, \citet{Fischer2017} measure a powerful exploding granule expanding with a rate of about $0.77\arcsec{}^{2}$ per minute, corresponding to an evaluate divergence amplitude of $5.5\cdot 10^{-4}\:\mathrm{s^{-1}}$.
In our case, the de-convolved HMI images have a pixel of 0.25\arcsec, which is about 183\:km. To compute the divergence, we used a spatial window of 0.9\:Mm (1.25\arcsec) and a temporal window of 3.75 min.
Our divergence amplitudes are in agreement with the values of \citet{Verma2011} and \citet{SBN94}  when our spatial and temporal windows are taken into account. The maximum of the divergence histogram of  the divergence amplitude inside the granule with an area greater or equal to 30 pixels is about $4\cdot 10^{-4}\:\mathrm{s^{-1}}$. We tested different thresholds above this value (see below).

Finally, for the vertical velocity we used a threshold of $0.9\:\mathrm{km\,s^{-1}}$. This
corresponds to the maximum of the histogram of the maximal Doppler value measured inside all the granules with an area greater or equal to $1\:\mathrm{Mm^{2}}$.

Table~\ref{tab:density} gives the sensibility of the measured exploding granule density relative to the choice of the area and divergence amplitude. We do not show the sensitivity to the threshold on radial velocity, since we test in Sect.~\ref{ssec:noDop} the impact of relaxing this criterion. 
We note a larger sensitivity to the area threshold, but a reduced effect concerning the choice of the divergence amplitude threshold. These densities have to be compared to that found by \citet{Title1989}, which is  $2.94\cdot 10^{-5} \mathrm{Mm^{2}\,s^{-1}}$, representing 41 exploding granules during a short sequence of 1360 seconds over a small field of view of $32\arcsec\times32\arcsec$. We conclude that lowing the threshold on the granule area leads to excessive false detection.
 
\begin{table}
    \caption{Density of exploding granule density related to the area and divergence amplitude for a Doppler threshold of $0.9\:\mathrm{km\,s^{-1}}$.}
    \label{tab:density}
    \centering
    \begin{tabular}{c c c }
        \hline\hline 
        Area & 
        Divergence &
        Density \\
        $\mathrm{Mm^{2}}$ & 
        $\mathrm{s^{-1}}$ &
        $\mathrm{Mm^{2}\,s^{-1}}$\\
        \hline
         0.66 & 5.47 $10^{-4}$  & 1.63 $10^{-4}$  \\
         0.66 & 6.56 $10^{-4}$  & 1.52 $10^{-4}$  \\
         1.00 & 5.47 $10^{-4}$  & 7.90 $10^{-5}$  \\
         1.00 & 6.56 $10^{-4}$  & 7.46 $10^{-5}$  \\
        \hline
    \end{tabular}
\end{table}

\begin{figure}[!htbp]
\includegraphics[width=\linewidth]{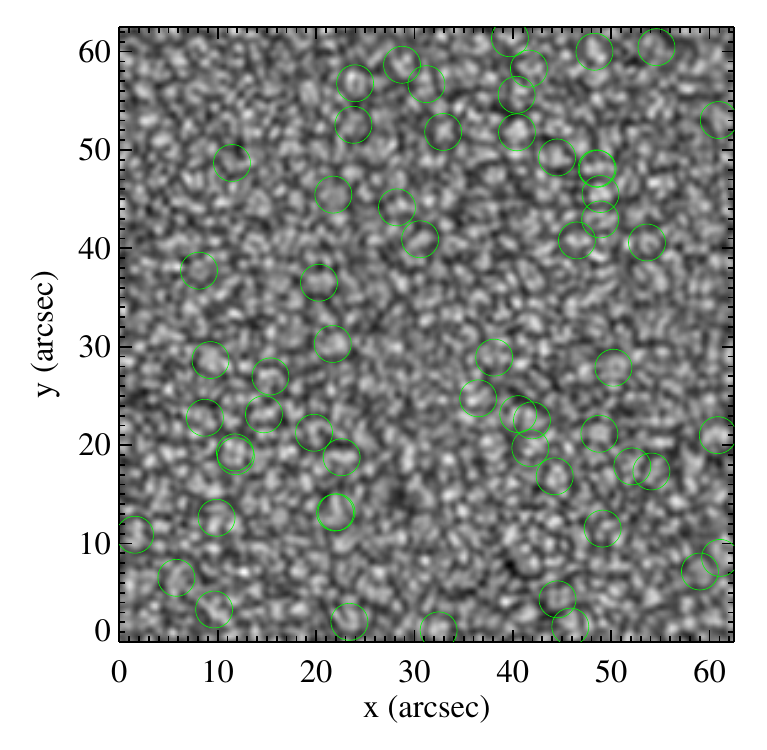}
\caption{Example of an SDO/HMI white-light intensity image of the granulation where automatically detected exploding granules are indicated by green circles, which allows visual control. The field of view is located at the disc centre of size $62.5\arcsec\times62.5\arcsec$.}
\label{granexp_1}
\end{figure}

We then chose as the criteria to automatically detect exploding granules:
\begin{enumerate}
    \item an area just before the explosion (that is, before the separation into several pieces) larger than $1\:\mathrm{Mm^{2}}$ (30 pixels),
    \item a divergence greater than $6.56 \cdot 10^{-4}\:\mathrm{s^{-1}}$, representative of exploding granules,
    \item a threshold on the ascent velocity of $0.9\:\mathrm{Mm^{2}\,s^{-1}}$.
\end{enumerate}
We then made a visual check of the automated detection of the exploding granules in several images. Figure~\ref{granexp_1}  gives an example of the automated detection.

\subsection{Exploding granules detected at the disc centre and their diffusion properties}
\label{ssec:diff}
This procedure allows for the detection and selection of exploding granules at the disc centre on the 24h sequence described in Sect.~\ref{sec:red}.  We aim to verify the ability of detected exploding granules to diffuse the magnetic field of the quiet Sun towards the photospheric network. To do so, we studied how passive test particles
(corks) are advected by the selected exploding granules only, without large-scale motions present in the  observed field. 
We used a technique described for example in \citet{SW89} or \citet{Rieutord2000}. It consists of integrating the trajectories of floating corks, which are initially uniformly distributed, and passively advected by a horizontal velocity field. We then analyse their final distribution. In the following, we describe the velocity field we used to simulate the impact of exploding granules alone.

We simulated each detected exploding granule with an isotropic function in the two dimensions $(x, y)$ data cube around its explosion time, $t$.
More precisely, we modelled the horizontal velocity inside an exploding granule during its lifetime as a divergent flow around its centre with a modulus,
 \begin{equation}
     u_h(r)= u_0 \frac{r}{R} \exp \left[-\frac{1}{2}\left(\frac{r}{R}\right)^2\right],
 \end{equation}
 where $r$ is the distance to the centre, $R$ the radius of the exploding granule, and $u_0$ the amplitude of the field velocity. This profile was obtained by assuming that the source of the horizontal flow follows a Gaussian distribution. This simple kinematic description of granules has already been used and discussed by \citet{Roudier2023}.
 
 \begin{figure*}[!htb]
\centering
\includegraphics[width=0.8\linewidth]{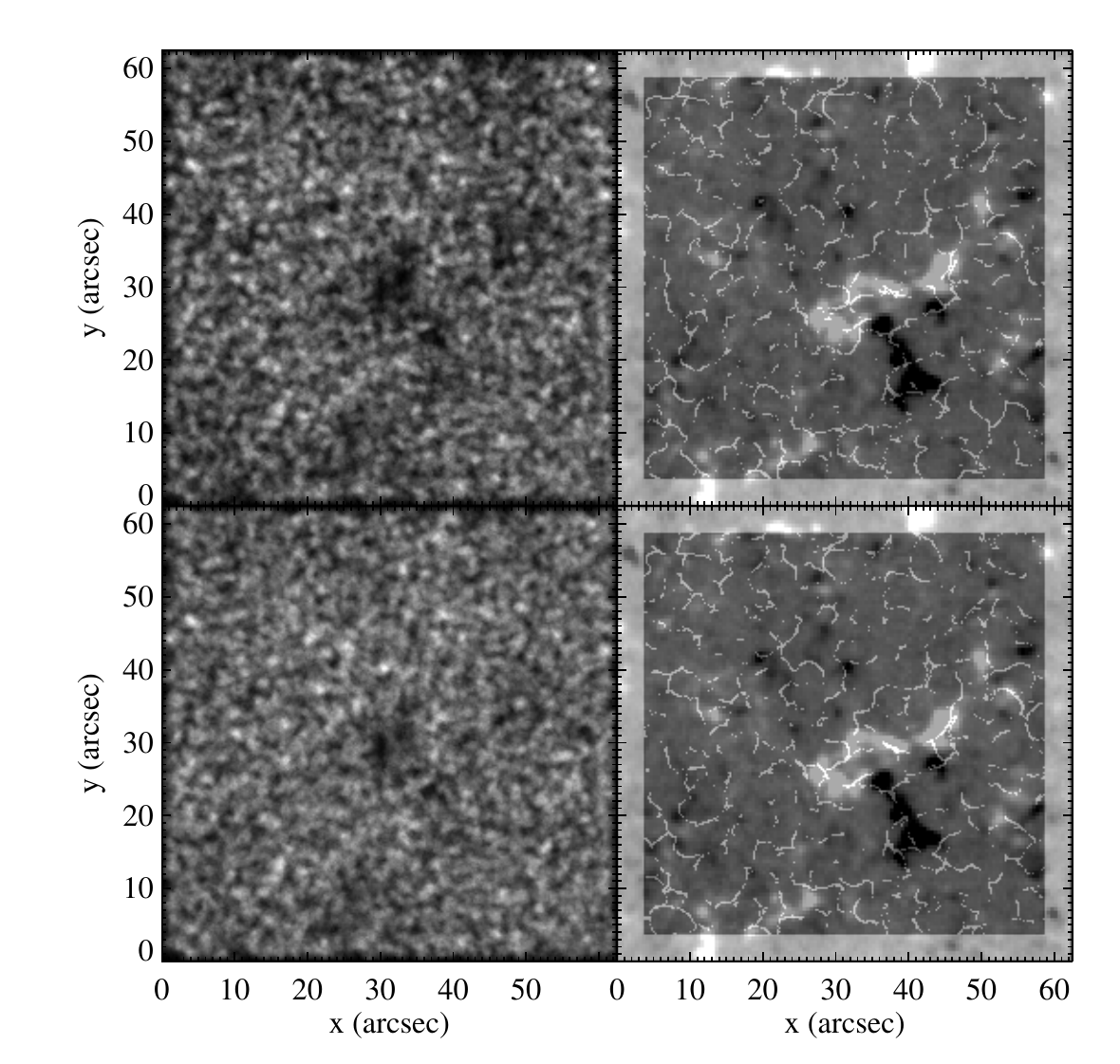}
\caption{Detected exploding granules during 24~h and their impact on corks. \emph{Left column:} Intensity maps constructing by summing up all exploding granules detected during 24~h: bright and dark zones indicate high and low rates of exploding granules, respectively. \emph{Right column:} Location of corks diffused by these detected exploding granules using our kinematic model, superimposed over maps of the longitudinal magnetic field observed by SDO/HMI at the end of the 24-h run. The exploding granule detection was performed with a criterion for the vertical velocity \emph{(top row)} and without \emph{(bottom row)}.
The field of view is located at the disc centre of size $62.5\arcsec\times62.5\arcsec$.
}
\label{granexp_comp}
\end{figure*}

 The radius, $R$, is proportional to the size of each exploding granule.
  The amplitude,  $u_0$,  was fixed to $2\:\mathrm{km\,s^{-1}}$, which corresponds to the initial expansion speed of exploding granules \citep{Roudier2020}. The lifetime of an exploding granule is not a well-defined quantity, because it is difficult to precisely define its beginning, and even more its end. However, for our diffusion simulations, we fixed 
  the lifetime to 3~min, which corresponds to the most active phase of this kind of granule, as is described in \citet{Roudier2020}. 
  This choice probably underestimates
  the duration of action of the exploding granules, but with this restriction we aim to evaluate the influence of this type of granules, alone, on its environment and the diffusion of the magnetic field on the solar surface.

We built a map that localises the position of detected exploding granules. In practice, for each detected exploding granule we built an image full of zeros, except for the pixels occupied by the exploding granule, which were set to one (we considered the pixels occupied by the exploding granule when it reaches its maximal area). Afterwards, we just summed all the images. The resulting map is shown on the top row of Fig.~\ref{granexp_comp} in the left panel. The location of corks at the end of the 24-h simulation is shown in the right panel. It is superimposed with a map of the longitudinal magnetic field component observed by SDO/HMI at the same time.
First, we observe a lower density of exploding granules where the magnetic field is higher, which corresponds to the photospheric network. This is in accordance with previous results that indicate a reduction of convective motions in magnetic regions, and therefore fewer exploding granules \citep{Title1989}.
Second, we observe that corks tend to gather where the magnetic field is higher, in particular in the photospheric network. This indicates that the most powerful exploding granules are among the main contributors to the construction of the photospheric network.

We also applied our automated pipeline to the data from one of our previous works \citep[][Fig.~3]{Roudier2020}. The automatic method detected 25\% of the exploding granules that were manually selected in this previous study. As was expected, we lost the ones with the smallest area, but we correctly detected the most dynamic and impactful granules in terms of surface occupation.

 \subsection{Detecting exploding granules without Doppler velocity}\label{ssec:noDop}

While the Doppler velocity is a measurement of the vertical velocity field at the disc centre, this is not the case at higher latitudes, where the Doppler signal results from both vertical and horizontal velocities. Therefore, the detection of exploding granules at higher latitudes or different longitudes cannot rely on a criterion based on Doppler velocity.

In this context, we want to test whether we can get rid of our third selection criterion. To do so,
we applied the same processing chain by applying thresholds only on the granule area and the velocity field divergence.
The bottom row of Fig.~\ref{granexp_comp} shows the results of the detection with these two selection criteria alone.
Certain exploding granules (one third) are detected in consecutive images. By means of three-dimensional recognition, we eliminated redundant detections. 
As we can see in Fig.~\ref{granexp_comp} (bottom), the action of the simulated exploding granules diffuses the corks to roughly the same locations, apart from a few details, after 24 hours.

In conclusion, the threshold on radial velocity field can be omitted when analysing exploding granules at different latitudes, by keeping only criteria on area and divergence. Thus, we can also track exploding granules even when we only have access to intensity images. We shall explore this possibility in the two next sections.

\section{Exploding granules relative to trees of fragmenting granules}\label{sec:TFG}

The location of exploding granules relative to large-scale convective structures
is also important for understanding the evolution of these convective scales and of magnetic field diffusion in the quiet Sun.
 \citet{Title1989} showed that exploding granules tend to occur inside mesogranules in regions of positive divergence of the horizontal flow. They concluded that almost all the granules in the centres of mesogranulation cells `either explode, are pushed aside, or are terminated by an exploding granule in their vicinity' (verbatim).
They also found that the granules born in regions of convergence (that is, negative divergence) are constrained from exploding.

To explore how exploding granules relate to mesogranulation cells, we have studied how they relate to TFGs, since the different branches of TFGs correspond to the mesogranular scale \citep{RRBRM2009}.
To locate exploding granules relative to TFGs, we cannot use SDO/HMI data, since its spatial resolution is not high enough. However, we already had a 48h sequence of granulation obtained by \textit{Hinode} from 29 to 31 August 2007, where TFGs were detected at high spatial resolution by \citet{Roudier2016}, following the method described in \citet{Roudier2003}.

Since we want to apply our automated exploding granule detection method, we needed, as a first step, to compare the results obtained with \textit{Hinode} and SDO/HMI in a simultaneous sequence. To do this comparison, we had access to a 1h55m-long sequence available on the same day and at the same hours (30 August 2010) at the disc centre with both instruments. 
We first reduced the resolution of \textit{Hinode} images from 0.11\arcsec\ to the SDO/HMI resolution (0.25\arcsec). We used on the \textit{Hinode} images the same automated detection method that is described in Sect.~\ref{sec:detect_meth}, except we needed to modify the threshold for divergence to $5.47\cdot 10^{-4}\:\mathrm{s^{-1}}$ to account for the different temporal sampling of the two sequences (50.2~s for \textit{Hinode} instead of 45~s for HMI).
 Despite the adjustment of this criterion, we notice that we detect slightly fewer powerful exploding granules in the \textit{Hinode} data, but the difference is very small.

 \begin{figure}[!htbp]
 \centering
\includegraphics[width=\linewidth]{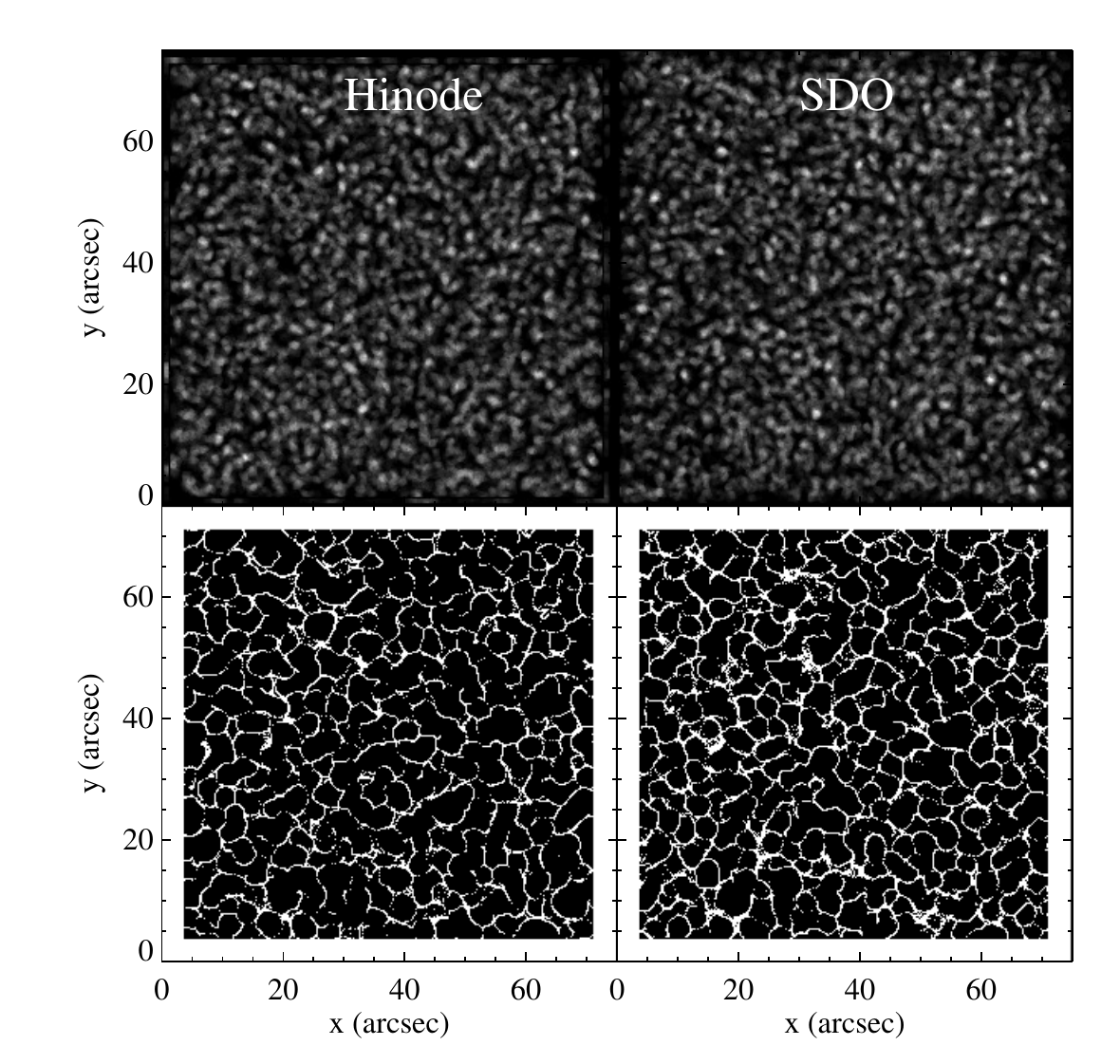}
\caption{Comparison of exploding granules detected with \textit{Hinode} and SDO/HMI in a simultaneous sequence. \emph{Top:} Intensity maps constructing by summing up  all exploding granules detected during 1h55m on \textit{Hinode} \emph{(left)} and SDO/HMI \emph{(right)} with a pixel of 0.25\arcsec. 
\emph{Bottom:}  Location of corks diffused by these
detected exploding granules using our kinematic model
using \textit{Hinode} \emph{(left)} and SDO/HMI \emph{(right)}. The field of view covers $75\arcsec\times75\arcsec$ at the disc centre.}
\label{granexp_2}
\end{figure}

 \begin{figure}[!htbp]
 \centering
\includegraphics[width=0.8\linewidth]{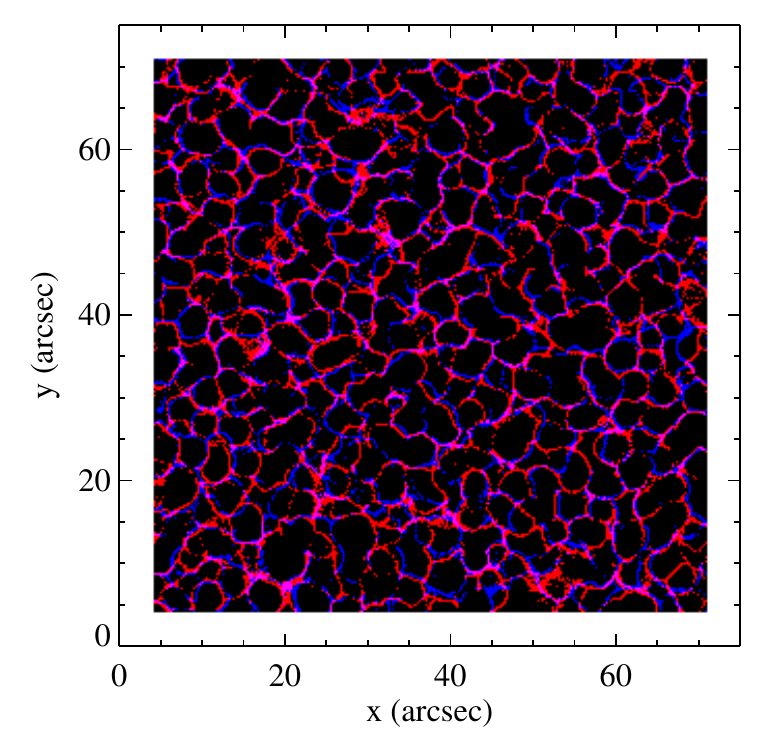}
\caption{Comparison of the cork locations diffused by the exploding granules detected in \textit{Hinode} and SDO/HMI (see Fig.~\ref{granexp_2}). Blue and red pixels show the locations of corks using \textit{Hinode} and SDO/HMI, respectively. Magenta indicates the superimposition of both. The field of view covers $75\arcsec\times75\arcsec$ at the disc centre.}
\label{granexp_2_comp}
\end{figure}

Figure~\ref{granexp_2} (top) shows the comparison of the total exploding granules detected with \textit{Hinode} and SDO/HMI during the full 1h55m-long sequences. 
Both maps are very similar, indicating that almost the same exploding granules have been detected. We only note some amplitude differences for some of them.
Nevertheless, the diffusion of the corks by the simulated exploding granules of the two sequences and their final positions reveal strong similarities (Fig.~\ref{granexp_2}, bottom). 
To highlight these similarities, we overlaid both maps in Fig.~\ref{granexp_2_comp} with two different colours, red and blue. A large number of the corks appear in magenta (that is, both colours). This indicates that, if both maps are not perfectly identical, they are quite similar: the corks generally surround the same cells, even if we notice that some cell boundaries are slightly shifted or incomplete.
This test confirms that we can use \textit{Hinode} data at a spatial resolution of $0.25\arcsec$ to automatically detect exploding granules.
Once validated on this short sequence, we applied the method to the full sequence for which TFGs have already been detected.

\begin{figure*}[!htbp]
\includegraphics[width=\linewidth]{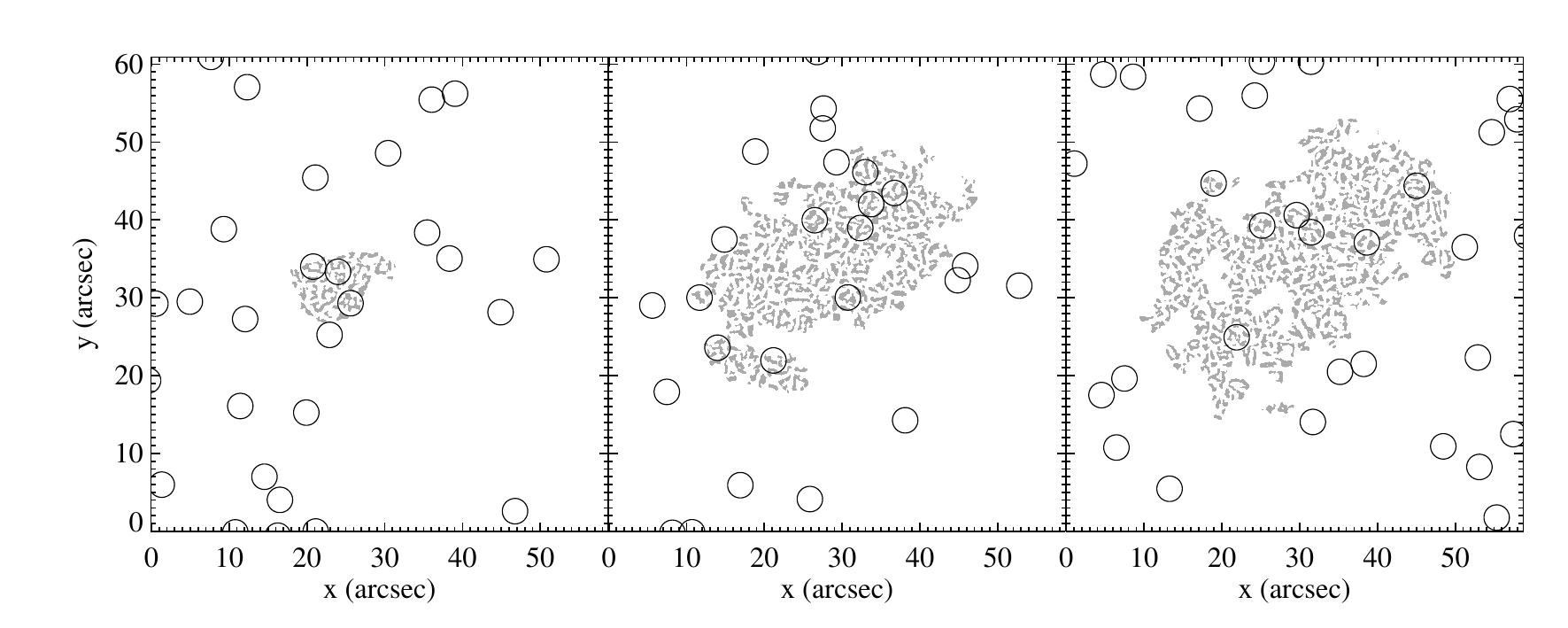}
\caption{Location of exploding granules (circle) relative to the TFG (in grey). Each panel corresponds to a different time: 
102 min \emph{(left)}, 441 min \emph{(middle)}, and 513 min \emph{(right)} after the birth of the TFG.
The field of view covers $59.3\arcsec\times61.5\arcsec$ at the disc centre.}
\label{granexp_4}
\end{figure*}

\begin{figure}[!htbp]
\includegraphics[width=\linewidth]{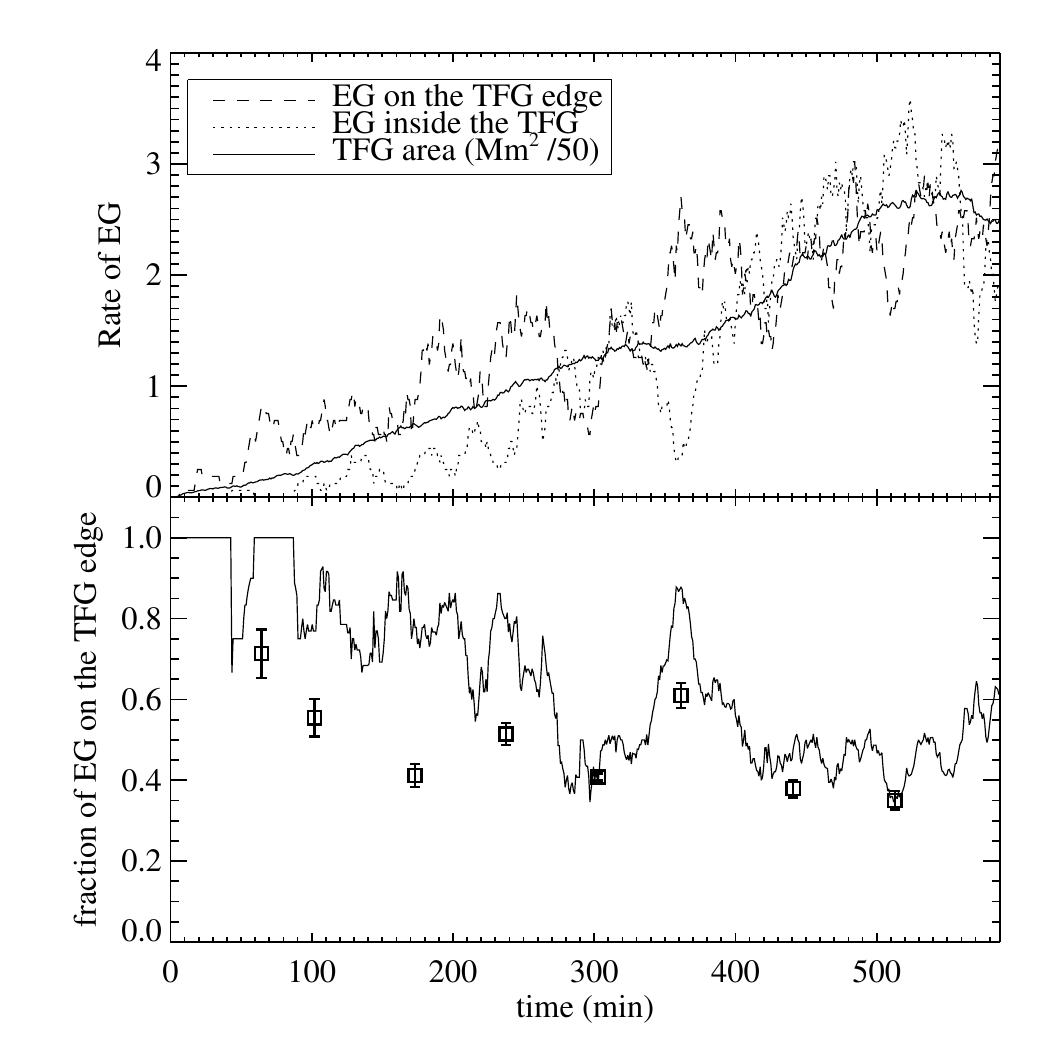}
\caption{Evolution of exploding granules in the TFG studied in this work. \emph{Top:} Number of exploding granules detected per minute on the edge of the TFG (dashed line) and inside the TFG (dotted line). The curves were smoothed with a 15-min-wide boxcar. For comparison, the evolution of the TFG area is also plotted (solid line, expressed in Mm$^{2}$ divided by a factor of 50). The origin of time is the birth of the TFG. \emph{Bottom:} Fraction of exploding granules of the TFG appearing on the edge (solid line). Squares show the fraction of granules of the TFG lying on its edge at some time steps (64, 102, 173, 238, 303, 361, 441, and 513~min). Error bars indicate the standard deviation obtained by counting the granules in consecutive images (see text for details).
}
\label{granexp_3}
\end{figure}

Once detected, we located exploding granules relative to one of the TFGs identified in \citet{Roudier2016}. More specifically, we chose the one shown in a grey colour in Fig.~A.1 Panel 3 of that article.
We identified exploding granules that are located inside the TFG or on its edge for 10 hours.
This identification was carried out manually. The exploding granules of the TFG surrounded by other granules of the TFG are considered to be inside the TFG; the others are located on its edge.
Figure~\ref{granexp_4} displays some examples of the location of the exploding granules (circles) relative to this TFG at different times: 102, 441, and 513 min from the birth of the TFG.

Figure~\ref{granexp_3} (upper panel) shows how the numbers of exploding granules inside the TFG and on its edge evolve. 
We have also plotted the evolution of the TFG area as a comparison reference, since, quite obviously, a wider TFG means more granules. 
From these measurements, we also deduced the fraction of exploding granules of the TFG located on its edge, defined as the ratio between the number of exploding granules on the edge and the total number of exploding granules in the TFG -- including those on its edge (Fig.~\ref{granexp_3}, lower panel).
This fraction decreases slowly (staying around 80\% for more than 200~min) before finally oscillating between 40 and 60\%. Qualitatively, such behaviour would be expected even if exploding granules appeared randomly. Indeed, just after the birth of a TFG all of the granules are on its edge, then the ratio between the boundary and the total area decreases (with the noticeable exception of the burst visible around $t=360$~min due to the twisted shape of the TFG at this moment, which temporally increases the length of the boundary). To verify whether the appearance of exploding granules is random or not, we also counted at several times the number of granules (including all kinds of granules, exploding or not) on the edge of the TFG and divided it by the total number of granules of the TFG. We performed this counting at $t= 64$, 102, 173, 238, 303, 361, 441, and 513 min (see square symbols on Fig.~\ref{granexp_3}, lower panel). To limit the impact of statistical fluctuations, we performed the manual count on 11 consecutive images around each considered time, $t$. 
We plot the standard deviations as error bars. We thus notice that the fraction of exploding granules on the edge is higher than the fraction obtained by considering all the granules. This  is especially noticeable during the initial growing phase of the TFG (before $\sim$300 min). We may assume that the highest occurrence of exploding granules at its edge during the initial phase helps the TFG to grow and develop. This is consistent with previous works. For example, \citet{Roudier2016} show that the locations of maxima of horizontal velocities coincide with the edges of nascent TFGs. Such high velocities thus appear to be directly related to exploding granules. In later phases of the TFG evolution, we notice that the fraction of exploding granules on the edge becomes similar to the one obtained by considering all the granules. This indicates that the edge is no longer a preferred area for exploding granules to appear.

\section{Exploding granules at different latitudes}\label{sec:lat}

One of the advantages of SDO/HMI observations is the complete coverage of the solar disc. This gives us access to the detection of exploding granules at various positions on the solar surface. We thus want to search for a possible variation with latitude of the number of exploding granules. To take into account various projection effects that would affect the measurements, we proposed comparing south-north variations (along a meridian) with measurements done in the east-west direction (along the equator), which we would take as a reference. Given the large volume of data to be processed, we limited our analysis to a shortened sequence of 5.2 hours.

We selected several fields at different latitudes (longitudes) along the central meridian (the equator): $-60\degr$, $-45\degr$, $-30\degr$, $-15\degr$, $0\degr$, $15\degr$, $30\degr$, $45\degr$, and $60\degr$.
For each position, we extracted a field of $250 \times 250$ pixels -- that is, $(45.7)^{2}\:\mathrm{Mm^{2}}$ -- at the centre of the disc. 

To apply our criterion to the granule area (area > $1\:\mathrm{Mm^{2}}$), we needed to compute them by taking into account the projection effect, by dividing the apparent area by the cosine of the angular distance between the field and the disc centre. However, when approaching the solar limb, small granules tend to cluster together, resulting in biased results. Hopefully, this project effect is the same in the south-north and east-west directions; thus, we can compare the longitudinal and latitudinal profiles.

We then measured the projection of the photospheric velocity field ($u_x$, $u_y$) onto the plane of the sky by using LCT. As was previously described, the velocities, $u_x$ and $u_y$ (in km/s), were computed at a cadence of 3.75 min with a spatial window of five pixels, equivalent to 1.25\arcsec. The combination of the results with the Doppler velocity maps, which provide the out-of-plane component, $u_z$, of the velocity field, allowed us to calculate the full vector velocity field at the surface in solar spherical co-ordinates, $u_{r}$,  $u_{\theta}$, and $u_{\phi}$, where $\theta$ and $\phi$ denote the co-latitude and the longitude, respectively \citep[see][for details]{ROUD2018}. 
The divergences were calculated from $u_{\phi}$ and $u_{\theta}$ in spherical co-ordinates. We applied our criterion (div > $6.56 \cdot 10^{-4}\:\mathrm{s^{-1}}$) to this field.

It is known that using LCT over the whole Sun produces shrinking effects \citep{Lop2016}. Since only local derivatives in latitude and longitude involve in the computing of divergence, we expect to be weakly impacted by this effect. Moreover, both the south-north and the east-west directions will once again be affected in the same way by a residual shrinking effect, since it only depends on the distance to the disc centre.

\begin{figure}[!htbp]
\includegraphics[width=\hsize]{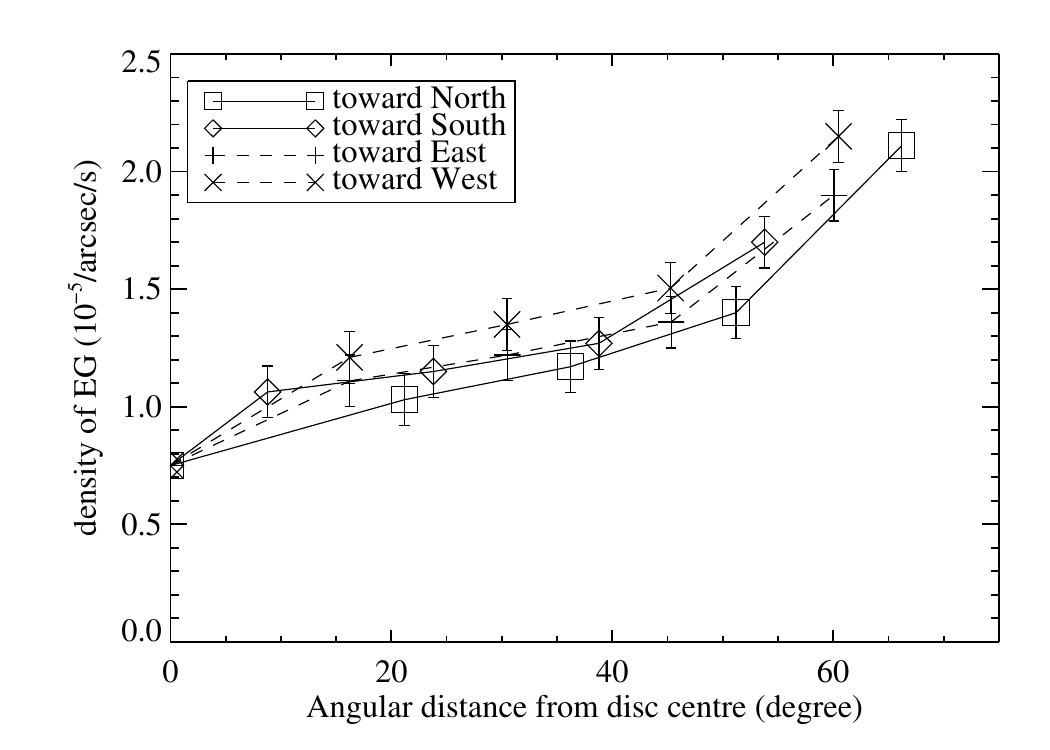}
\caption{Density of exploding granules detected at different latitudes and longitudes plotted as a function of the angular distance from disc centre. Solid lines join points along the central meridian in the northern (squares) and southern hemisphere (diamonds). Dashed lines join points along the equator in the eastern (plus signs) and western (crosses) directions.}
\label{granexp_5}
\end{figure}

We extracted the mean exploding granule density per Mm$^{2}$ and per second for the 17 fields we analysed. These results are plotted in Fig.~\ref{granexp_5} as a function of the angular distance from the disc centre, which is the suited parameter with which to characterise the projection effects. The same latitudes in the northern and southern hemispheres do not correspond to the same angular distance from the centre, because the $B_0$ angle was $-6.2\degr$.
Apart from measurement errors (computed as standard deviations), the behaviours are very similar for the four cardinal directions: all the curves follow the same trend.
We can thus conclude that the density of exploding granules does not significantly vary with latitude.

\begin{figure}[!htp]
\centering
\includegraphics[width=7.5cm]{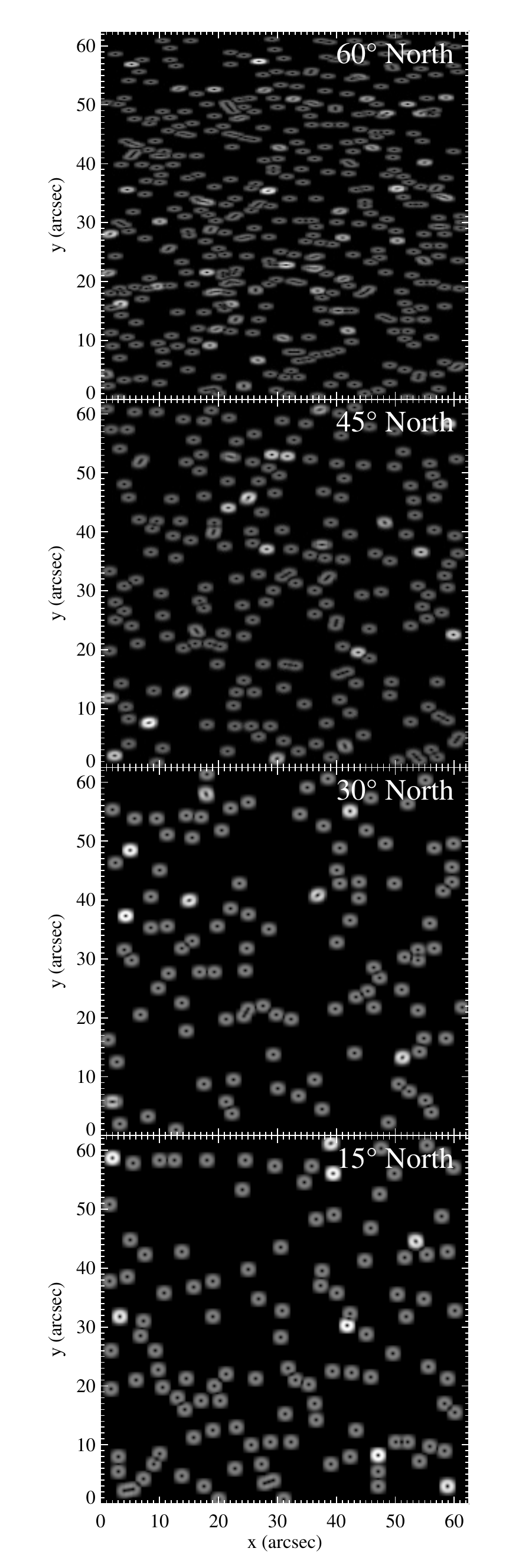}
\caption{Simulated velocity modulus, $u_h$, of the exploding granules at different latitudes in the northern hemisphere, at $t=75$ minutes of the sequence. The fields of view cover $62.5\arcsec\times62.5\arcsec$.}
\label{granexp_6}
\end{figure}

As was previously done at the disc centre with the full 24-h sequence, we applied our cork diffusion kinematic model to the 5.2-h sequence in the 17 fields at different latitudes and longitudes. We simulated the exploding granules with the same functions described in Sect.~\ref{ssec:diff}, but we used  $u_{\phi}$ and  $u_{\theta}$ (instead of $u_x$ and $u_y$) and we took into account the projection effect, which transforms a circle into an ellipse as a function of the centre angle.  Figure~\ref{granexp_6} illustrates this kinematic model for the four fields in the northern direction.

\begin{figure}[!htp]
\centering
\includegraphics[width=7.5cm]{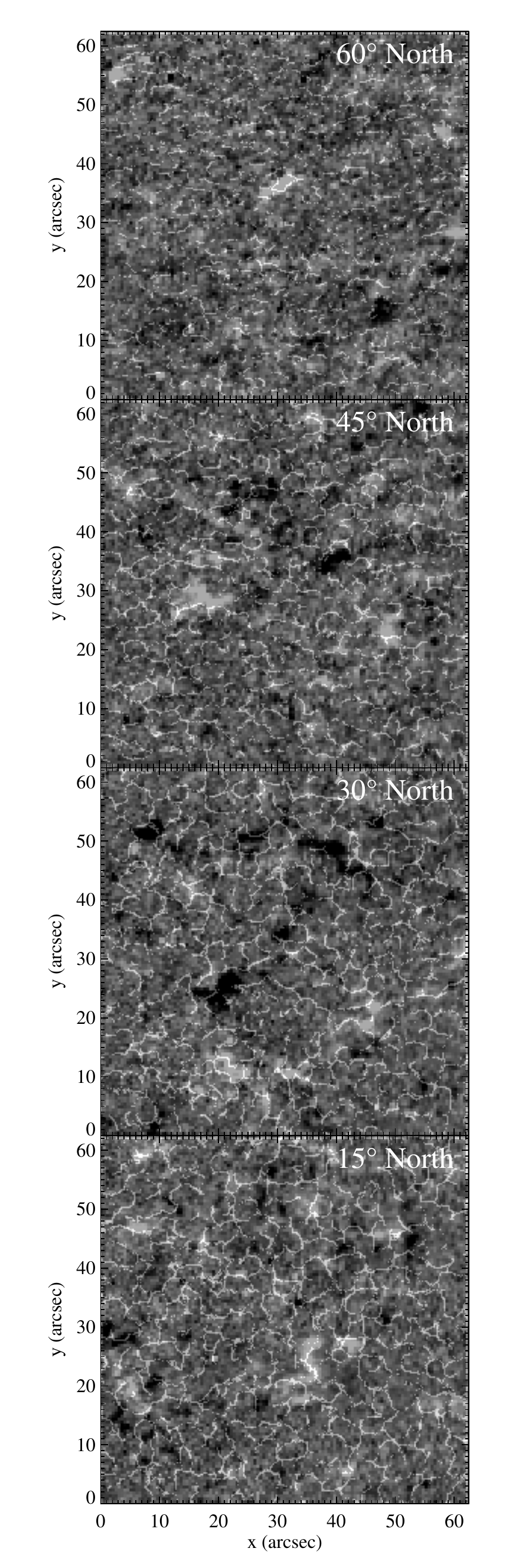}
\caption{Location of corks diffused by
detected exploding granules using our kinematic model, superimposed over maps the longitudinal magnetic observed by SDO/HMI after 5.2 hours. The fields of view cover $62.5\arcsec\times62.5\arcsec$.}
\label{granexp_7}
\end{figure}

Figure~\ref{granexp_7} shows the locations of corks,  superimposed over averaged magnetic maps at different latitudes in the northern hemisphere. In these figures, the exploding granules detected contribute to the strong scattering of the corks, but the duration of the processed sequence (5.2 hours) does not allow us to see the formation of the network.  Nevertheless, we notice some concentrations of corks on some network magnetic structures, even after such a short time, highlighting how the most intense exploding granules are important contributors to the diffusion of magnetic field at the solar surface.

\section{Discussion and conclusion}\label{sec:conc}

The whole solar surface is governed by turbulent convection.  The impact of exploding granules on the solar surface dynamics and magnetism has been studied by various authors in the last few decades. It has long been recognised that the few convective elements that are the most dynamic play a dominant role in magnetic field diffusion and also in acoustic wave generation. Indeed, according to \citet{Roth2010}, the exploding granules contribute to the excitation of solar p modes in addition to the contribution of intergranular downflow lanes.
The vigorous expansion of the exploding granules has radially outward-facing horizontal flows, causing, together with the horizontal flows from the surrounding granules, the magnetic elements in the bordering intergranular lanes to be squeezed and elongated. In reaction to the squeezing, \citet{Fischer2017} detect a chromospheric intensity and velocity oscillation pulse, which was identified as an upward travelling hot shock front, visible in \ion{Mg}{ii} h\&k lines. In addition, an enhancement in the intensity and velocity oscillations of high photospheric or chromospheric spectral lines is found in the region of the dark core of an exploding granule  \citep{ELL2021}. \citet{Gug2020} discussed the magnetic nature of exploding granules. They suggest that multipolar structures emerging into the photosphere, resembling an almost horizontal flux sheet, could be associated with exploding granules. Doppler shifts of up to 9 km/s are observed at the edges of bright granules, demonstrating that the flows reach supersonic speeds \citep{Bellot2009}. These supersonic flows may be linked to the exploding granules \citep{ObaPhD}.

In this paper, we have proposed a way to automatically detect the most dynamic convective elements, the exploding granules, in the SDO/HMI de-convolved images, freely available at JSOC,\footnote{\url{http://jsoc.stanford.edu}} by using their area and the amplitude of their divergences. At the disc centre, the Doppler velocity can be used as an additional criterion, but appears to be unnecessary. The method we propose here is quite simple to implement. It requires (i) a measurement of the granule area and (ii) a measurement of the horizontal velocity field. For the first measurement, a segmentation algorithm is needed. We have made publicly available such a tool in the Correlation Structure Tracking (CST) package.\footnote{The CST is available at the Multi Experiment Data \& Operation Center: \url{https://idoc.osups.universite-paris-saclay.fr/medoc/tools/cst-codes/}} For the latter measurement, the LCT is needed. Several implementations of the LCT have already been made public.\footnote{See for example \url{https://solarmuri.ssl.berkeley.edu/overview/publicdownloads/software.html} in IDL or \url{https://github.com/Cadair/pyflct} in Python.}

As a first step, we measured the density of exploding granules at the centre of the disc over a 24-hour period. We found a deficit of these dynamic elements where the magnetic field is stronger and more concentrated.
By modelling the horizontal velocity inside an exploding granule as a divergent flow around its centre, we simulated the action of all the detected exploding granules on corks. Looking at the resulting final corks positions after 24h of evolution, we noticed similarities between cork locations and the longitudinal magnetic field. Corks gathered where the magnetic field is higher, in particular in the network, even though we used a simple model, have only selected the most active divergent elements and have neglected the large-scale underlying motions. This illustrates how exploding granules are sufficient to form a large part of the photospheric magnetic network in the quiet Sun.

We also studied the position of exploding granules relatively to the TFGs. We noticed that the most intense exploding granules are preferentially located on the edges of a TFG, during a period of about $\sim 300$ minutes after the birth of the TFG.

We investigated a possible latitudinal dependency in the number of exploding granules. We find the same rate of exploding granules at different latitudes within the measurement errors, once we correct for the project effects.

Future works could focus on the evolution of the density of these exploding granules during the solar cycle, given the observed fluctuations in granulation properties at different phases of the cycle \citep{Ballot2021}. Other studies could also be carried out in the vicinity of active regions to determine, for example, the erosive effect of these exploding granules on sunspots.

\begin{acknowledgements}
We thank the anonymous referee for providing helpful and constructive comments.
We thank Jean-Marie Malherbe for very  useful discussions and advices.
The HMI data is courtesy of the SDO/HMI Science Investigation Team. Some computing was performed at the JSOC, which is located at the Stanford University and operated by the HMI Science Investigation Team. This work was granted access to the HPC resources of CALMIP under the allocation 2011-P1115. This work was supported by COFFIES, NASA Grant 80NSSC22M0162.
\end{acknowledgements}

\bibliographystyle{aa}    
\bibliography{biblio_eg}

\end{document}